\documentclass[review]{elsarticle}

\usepackage[a4paper, total={6.0in, 8.0in}]{geometry}
\usepackage{graphicx}
\usepackage{arydshln}
\usepackage{multirow}
\usepackage{xcolor}
\usepackage[flushleft]{threeparttable}
\usepackage[colorlinks = true, linkcolor = blue, urlcolor  = blue, citecolor = blue, anchorcolor = blue]{hyperref}
\usepackage{fixltx2e}

\graphicspath{ {./} }
\journal{Journal of Molecular Graphics and Modelling}

\begin{document}

\begin{frontmatter}

\title{OpenMX Viewer: A web-based crystalline and molecular graphical user interface program}

\author{Yung-Ting Lee}
\ead{ytlee@issp.u-tokyo.ac.jp}
\author{Taisuke Ozaki\corref{mycorrespondingauthor}}
\address{Institute for Solid State Physics, The University of Tokyo,\\5-1-5 Kashiwanoha, Kashiwa, Chiba 277-8581, Japan}
\ead{t-ozaki@issp.u-tokyo.ac.jp}

\begin{abstract}
The OpenMX Viewer (Open source package for Material eXplorer Viewer) is a web-based graphical user interface (GUI) program for visualization and analysis of crystalline and molecular structures and 3D grid data in the Gaussian cube format such as electron density and molecular orbitals. The web-based GUI program enables us to quickly visualize crystalline and molecular structures by dragging and dropping XYZ, CIF, or OpenMX input/output files, and analyze static/dynamic structural properties conveniently in a web browser. Several basic functionalities such as analysis of Mulliken charges, molecular dynamics, geometry optimization and band structure are included. In addition, based on marching cubes, marching tetrahedra and surface nets algorithms with Affine transformation, 3D isosurface techniques are supported to visualize electron density and molecular/crystalline orbitals in the cube format with superposition of a crystalline or molecular structure. Furthermore, the Band Structure Viewer is implemented for showing a band structure in a web browser. By accessing the website of the OpenMX Viewer, the latest OpenMX Viewer is always available for users to visualize various structures and analyze their properties without installations, upgrades, updates, registration, sign-in and terminal commands.
\end{abstract}


\end{frontmatter}

%\linenumbers

\section{Introduction}

Recently, a number of theoretical and computational research studies based on first-principles electronic structure calculations have grown in a variety of scientific fields \cite{ref1,ref2,ref3,ref4}. To analyze complicated data obtained by density functional theory (DFT) or other first-principles approaches, many visualization software such as XCrySDen \cite{XCrySDen1,XCrySDen2}, VMD \cite{VMD}, VESTA \cite{VESTA} and Mercury \cite{Mercury1,Mercury2}, have provided platforms to visualize various crystalline or molecular structures and to analyze data from simulations to explain static and dynamic properties of materials. These visualization software are useful and important to reduce time for building or modifying structures, analyzing complicated calculation results, and measuring structural information. The use of such software enhances the efficiency of dealing with the input/output data for calculations and analyzing simulation results.
\\
Since its release in 2008, HTML5 has introduced application programming interfaces (APIs) that can be connected with Javascript for constructing web applications in web browsers \cite{3DGraphicsOnTheWeb}. Also, Web Graphics Library (WebGL) (\url{https://www.khronos.org/webgl/}) is a Javascript-based API developed for interactive 3D graphical objects without plug-in in web browsers, which is able to accelerate image processing by GPU for web canvas. With the advent of WebGL, Three.js, a cross-browser Javascript library, is able to utilize WebGL rendering techniques and to display animated 3D computer graphics in modern web browsers \cite{ThreeJS}. Based on HTML5 and WebGL, one can build an application in a web browser with the same functions as one installed in conventional operating systems. There are several well-established examples of web-based viewers for chemistry \cite{NGLref9}, biochemistry \cite{NGLref19} and medical imaging \cite{MedicalImaging} such as: JSmol \cite{JSmol}; ChemDoodle \cite{ChemDoodle}; Chemozart \cite{Chemozart}; Web3DMol \cite{Web3DMol}; 3Dmol.js \cite{3Dmol}; LiteMol \cite{LiteMol}; and, NGL Viewer \cite{NGLViewer}. Therefore, it will be expected that a web application can be developed to visualize crystalline and molecular structures which may provide versatile tools for analyzing simulation data. Furthermore, it is possible to use the same functions as applications in a web browser when users are offline.
\\
The OpenMX Viewer is an interactive web-based crystalline and molecular structure viewer based on HTML5 Canvas 2D combined with Javascript and Three.js libraries for existing or potential users of OpenMX \cite{OpenMX1}: (1) to visualize crystal structures, isosurfaces, and band strucutres; (2) to analyze static/dynamic data of OpenMX; and, (3) to save structural information in a XYZ format, a CIF format, or an OpenMX input format by using our unique javascript libraries and functions. It contains three kinds of viewers - Structure Viewer, Cube 3D Viewer and Band Structure Viewer. In particular, we develop a new on-line 3D-isosurface viewer for lattice systems, i.e. Cube 3D Viewer, by using the Three.js library, isosurface algorithms, our parsing Gaussian cube file library and a transformation of lattice vectors to visualize an isosurface image in a crystal lattice in a web browser. The OpenMX Viewer provides a convenient GUI interface to examine a structure or simulation data by dragging-and-dropping a file in a compatible web browser, such as Chrome, Firefox or Safari, without any installations, upgrades, updates, registration, sign-in or terminal commands. In addition to OpenMX input/output formats, XYZ, CIF, and Gaussian cube formats are accepted as an input file by the OpenMX Viewer with our parsing file libraries in javascript. The latest version of the OpenMX Viewer is always on the Internet for users to use freely anywhere and at any time. It is also noted that the OpenMX Viewer has been released under the GNU General Public License. Furthermore, the OpenMX Viewer has been developed not only for analyzing OpenMX input/output files but also for supporting XYZ, CIF, and cube file formats as an input file to visualize structures and electron density, enabling a wide variety of people using other software to obtain benefits with the OpenMX Viewer. This paper gives a brief overview of the OpenMX Viewer as below.

\section{The features of the OpenMX Viewer}

The OpenMX Viewer is built to connect with first-principle calculations for analyzing crystalline and molecular static/dynamic structures and relevant physical and chemical properties in the first and last step of simulations. Although the GUI has been originally developed for first-principles calculations by a DFT code, OpenMX, many functionalities explained later on can be utilized by users using other DFT codes because of the use of common file formats, such as XYZ, CIF, and cube. The OpenMX Viewer can be utilized directly by dragging-and-dropping files to the OpenMX Viewer webpage \cite{OpenMXViewer} without installations, upgrades, and terminal commands. The main features of the OpenMX Viewer include: (1) Structure Viewer; (2) Cube 3D Viewer; and, (3) Band Structure Viewer. It supports several kinds of data formats listed in Table \ref{table:t1}. After dragging and dropping a file to the OpenMX Viewer, it can automatically recognize a file type, read data and show results on the corresponding viewer. The Chrome and Firefox browsers are preferred for the OpenMX Viewer due to its good compatibility and performance. The main functionalities and implementations of the OpenMX Viewer are described in the following sections.

\begin{table}[h]
\caption{Viewers and corresponding supported file types} \label{table:t1} 
\begin{tabular}{lll}
\hline
\multicolumn{2}{c}{\textbf{Viewer}} & \multicolumn{1}{c}{\textbf{Supported file types}} \\ \hline
\multicolumn{1}{c}{\textbf{Structure Viewer}} & \multicolumn{1}{c}{(Static)} & XYZ type (.xyz), CIF type (.cif), OpenMX input (.dat) \\
\textbf{Structure Viewer} & (Dynamic) & XYZ type (.xyz), OpenMX output files (.md) \\
\multicolumn{2}{l}{\textbf{Cube 3D Viewer}} & Gaussian cube files (.cube) \\
\multicolumn{2}{l}{\textbf{Band Structure Viewer}} & Band files (.Band) \\ \hline
\end{tabular}
\end{table}

\section{The functionalities of the OpenMX Viewer}

\subsection{Structure Viewer}
The Structure Viewer is capable of reading a crystalline or molecular structure from standard file formats, such as XYZ files, CIF files \cite{CIF}, OpenMX input files (.dat) and OpenMX output files (.md), and displaying a structure by using drag-and-drop on the Structure Viewer to show static or dynamic structures in the orthographic or perspective view. It also includes analysis of Mulliken charges, spins, forces and velocities. Users can use the \textit{Analysis of Structure} panel to measure: (1) a bond distance between two selected atoms; (2) a bond angle; and, (3) a dihedral angle as shown in Fig. 1. If a dropped file is relevant to structural optimization or molecular dynamics simulations, users can select \textit{Structural Change} to check structural changes at each optimization or molecular dynamics step as shown in Fig. 2. In addition, after dragging and dropping a file to visualize the structure, users can save the structural information to another format, such as the XYZ format, CIF format or OpenMX input file format with Cartesian coordinates or fractional coordinates.

\subsection{Cube 3D Viewer}
The Cube 3D Viewer is built for showing 3D isosurface images in the Gaussian cube format \cite{Gaussian} such as electron density, difference electron density taken from superposition of atomic densities of constituent elements and molecular orbitals. To show an isosurface image, users can drag and drop a cube file to the OpenMX Structure Viewer window or the Cube 3D Viewer window as shown in Fig. 3. Atoms, a unit cell, and positive/negative isosurfaces can be rendered in the Cube 3D Viewer at the same time. Moreover, an isosurface image in a crystal structure can be extended to a larger supercell.

\subsection{Band Structure Viewer}
The Band Structure Viewer allows us to show and check a band structure quickly by dragging-and-dropping a Band file in the end of OpenMX's band structure calculation. It also includes a function to rescale a band structure by using mouse wheel or buttons.

\section{The implementations of the OpenMX Viewer}
\subsection{Structure Viewer}
The Structure Viewer is developed by HTML5 canvas 2D to display a static/dynamic crystalline or molecular structure in the orthographic or perspective view by dragging-and-dropping and parsing a XYZ/CIF/OpenMX file. After reading structural information, the spheres of atoms and bond stickers between two atoms are prepared and saved first. In the case of bond stickers, 91 bond stickers with tilted angles from 0 degrees to 90 degrees are drawn and stored in advance in order to reduce the effort of plotting bond stickers at each operation e.g. translations or rotations. Also, the images of spheres of atoms are created at the beginning of rendering operations. During the drawing of a crystal structure at each operation, these images of spheres and bond stickers are adjusted in size and pasted to their proper place. Then, the structure with spheres and bond stickers will be visualized on the center of the Structure Viewer. In Table \ref{table:t2}, we show the performance of the OpenMX Structure Viewer, indicating that the OpenMX Viewer provides a quick visualization of crystalline and molecular structures. In addition, the OpenMX Structure Viewer provides plenty of options listed in Table \ref{table:t3} for users to show structural properties. Here, eight practical options for examining structures are described as below:

\begin{enumerate}
\item A supercell can be chosen by 'Supercell' options directly. After selecting a supercell, the supercell structures will be changed and visualized simultaneously. 
\item 'Number' can be checked to show all of serial numbers on the top of spheres.
\item 'Symbol' stands for printing a symbol of atomic species on the top of each sphere.
\item When 'Structure' is checked, the '\textit{Analysis of Structure}' panel will be shown at the top-left corner of the Structure Viewer.
\item When 'Dynamics' is checked, the '\textit{Strucutral Change}' panel will be shown at the bottom of the Structure Viewer.
\item In the 'Spin' option, there are four cases: (1) off; (2) spin; (3) spin-C; and, (4) spin-V. If the second case is selected, numerical values of the spin of elements in units of $\mu_{B}$ will be printed on the top of spheres. And, '-C' represents the magnitude of spin by colors, while '-V' represents them by vectors.
\item For 'Force' and 'Velocity' options, both of them are shown by vectors pointing to a direction, while length of the vectors depends on their magnitude.
\item In the 'Save' option, there are four cases for saving the structural information: (1) 'xyz' is for saving a XYZ file; (2) 'cif' is for saving a CIF file; (3) 'OMX(xyz)' is for generating an OpenMX input file in a Cartesian coordinate system; and, (4) 'OMX(frac)' is for generating an OpenMX input file with fractional coordinates.
\end{enumerate}

\begin{figure}[h!]
	\centering
	\includegraphics[scale=0.7]{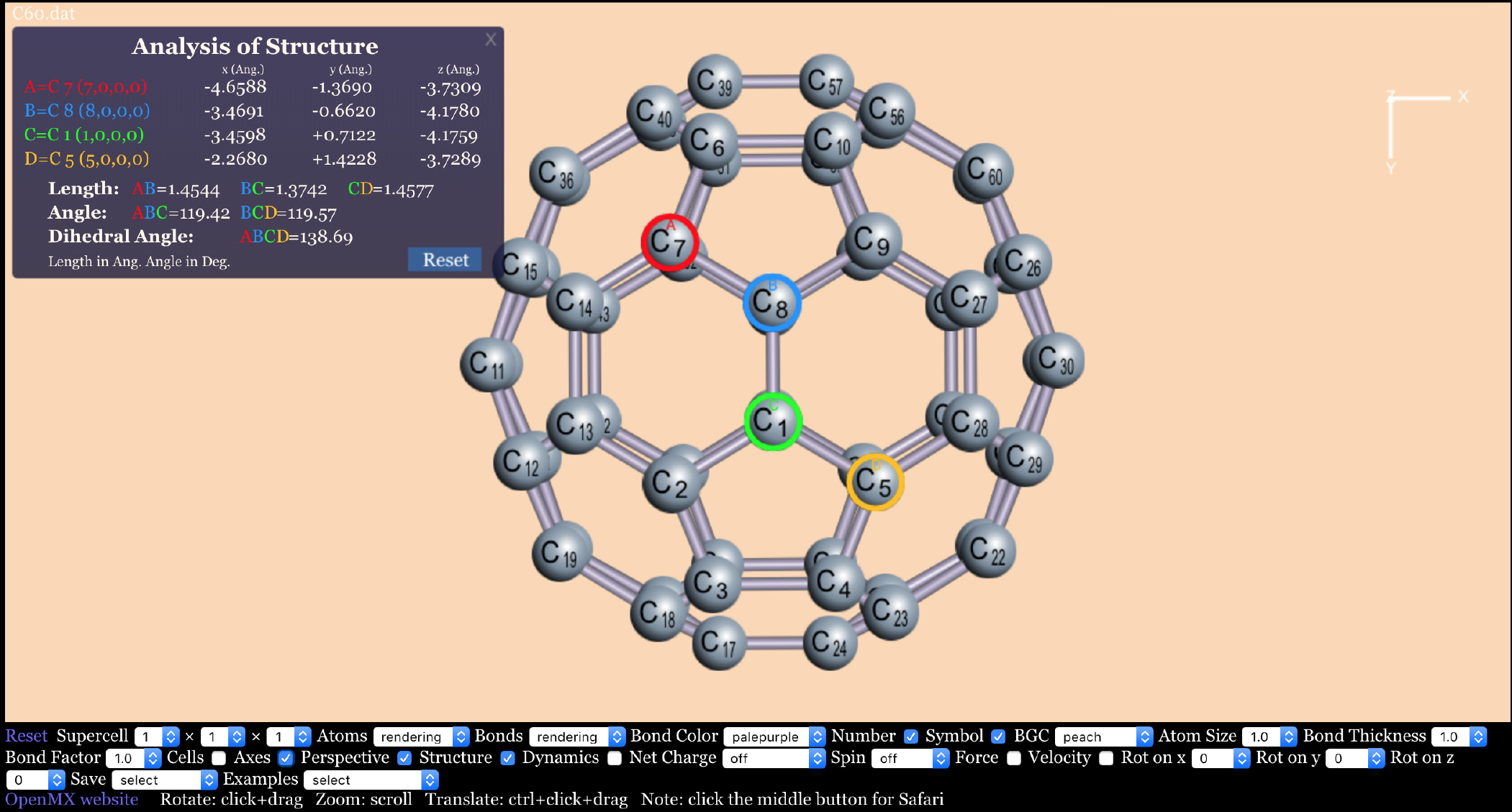} \label{figure:f1} 
	\caption{The OpenMX Structure Viewer with a C\textsubscript{60} structure in the perspective view. The panel of \textit{Analysis of Structure} is in the top-left corner. The options of structural properties and styles are listed at the bottom.}
\end{figure}

\begin{table}[h!]
\footnotesize
\caption{Performance of parsing and structure rendering a Si conventional cell with 3 different supercells.} \label{table:t2} 
\begin{tabular}{lcccrrrr}
\hline
\multicolumn{1}{c}{\multirow{2}{*}{System}} & \multicolumn{3}{c}{File parsing {[}s{]}} & \multicolumn{1}{c}{\multirow{2}{*}{\begin{tabular}[c]{@{}c@{}}Creating images\\ and rendering {[}s{]}\end{tabular}}} & \multicolumn{1}{c}{\multirow{2}{*}{\begin{tabular}[c]{@{}c@{}}Rotation and\\ rendering {[}s{]}\end{tabular}}} & \multicolumn{1}{c}{\multirow{2}{*}{\begin{tabular}[c]{@{}c@{}}Number\\ of atoms\end{tabular}}} & \multicolumn{1}{c}{\multirow{2}{*}{\begin{tabular}[c]{@{}c@{}}Javascript\\ Memory {[}MB{]}\end{tabular}}} \\
\multicolumn{1}{c}{} & XYZ & CIF & OpenMX & \multicolumn{1}{c}{} & \multicolumn{1}{c}{} & \multicolumn{1}{c}{} & \multicolumn{1}{c}{} \\ \hline
Si 5$\times$5$\times$5 supercell & 0.022 & 0.025 & 0.022 & 0.203 & 0.008 & 1000 & 11.703 \\
Si 10$\times$10$\times$10 supercell & 0.088 & 0.388 & 0.090 & 1.233 & 0.010 & 8000 & 29.703 \\
Si 15$\times$15$\times$15 supercell & 0.199 & 2.802 & 0.185 & 10.397 & 0.193 & 27000 & 59.297 \\ \hline
\end{tabular}
\begin{tablenotes}
  \footnotesize
  \item \textit{Notes:} Tests for the OpenMX Structure Viewer were performed in a Chrome browser in Linux Ubuntu 16.04 with a 4.2 GHz Intel Core i7-7700K Processor and 32 GB DDR4-2400 MHz Memory.
\end{tablenotes}
\end{table}

\subsubsection{Structural measurement}
The \textit{Analysis of Structure} panel is designed to show measurements of a bond distance between two selected atoms, a bond angle from three selected atoms and a dihedral angle formed by four selected atoms. After users select any four atoms in sequence, the panel will show the element symbols and their serial number and Cartesian coordinates as shown in Fig. 1. The three bond lengths between two selected atoms, two angles among three selected atoms, and one dihedral angle from four selected atoms are printed in the same panel.

\subsubsection{Animation panel}
The \textit{Structural Change} panel is capable of showing animation of optimizations or molecular dynamics simulations and printing total energy in Hartree at each step as shown in Fig. 2. Users can play an animation step-by-step or continuously from beginning to end and switch steps forward or backward. At the same time, selected bond lengths, bond angles, and a dihedral angle at each step are shown in the \textit{Analysis of Structure} panel.

\begin{figure}[h!]
	\centering
	\includegraphics[scale=0.7]{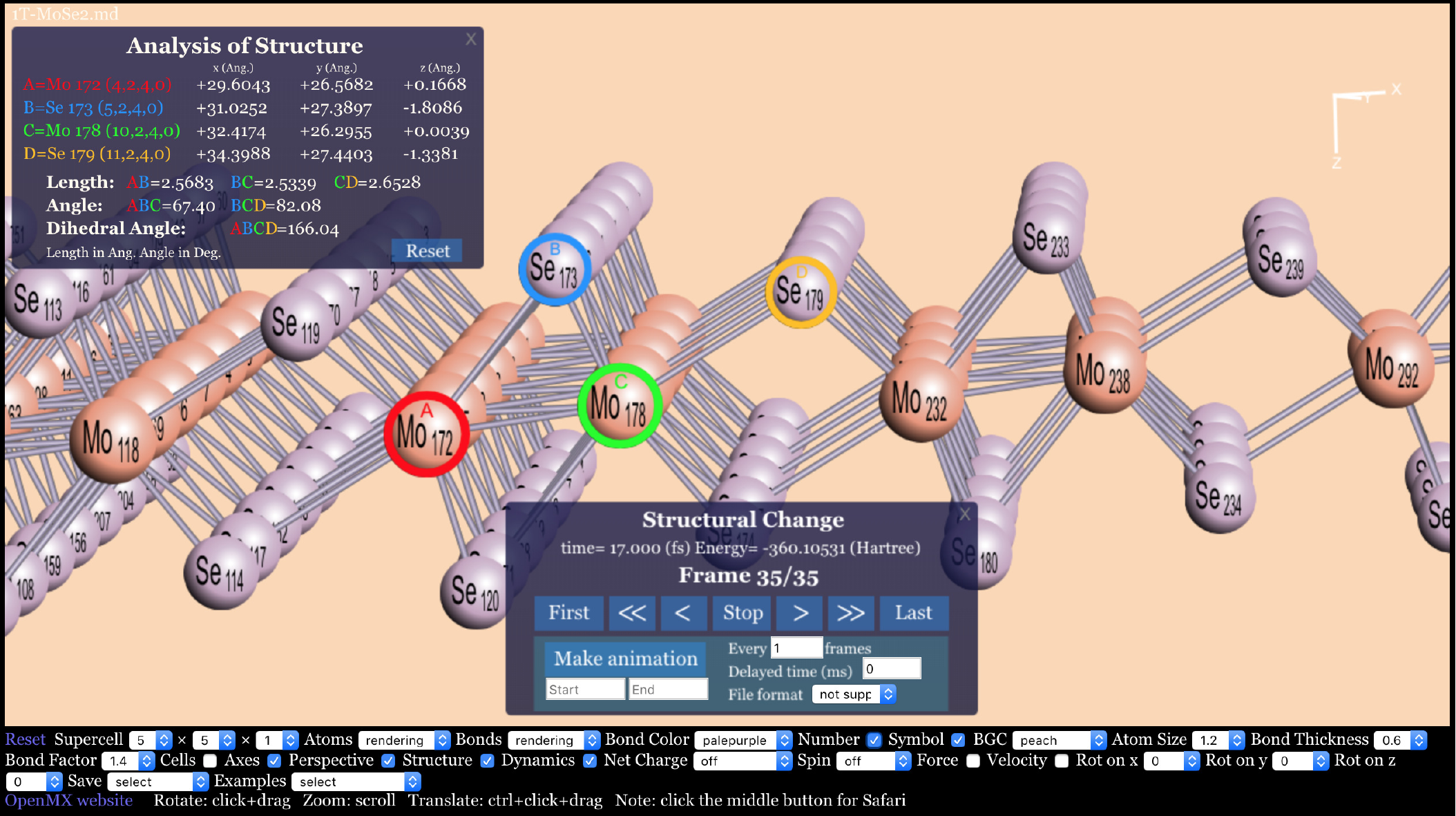} \label{figure:f2} 
	\caption{The OpenMX Structure Viewer with 1T-MoSe\textsubscript{2} structure in the perspective view. The \textit{Analysis of Structure} panel and the \textit{Structural Change} panel are in the top-left corner and the bottom, respectively.}
\end{figure}

\begin{table}[h!]
\caption{The key options of display styles and structural properties in the OpenMX Structure Viewer} \label{table:t3} 
\begin{tabular}{llc}
\hline
\multicolumn{1}{c}{\textbf{Options}} & \multicolumn{1}{c}{\textbf{Description}} & \multicolumn{1}{c}{\textbf{Default Value}} \\ \hline
Supercell & Setup the size of supercell ( a $\times$ b $\times$ c ) & 1 $\times$ 1 $\times$ 1 \\
Number & Show the serial number of atoms & off \\
Symbol & Show the element symbols & off \\
Bond Factor & Show bonds between two atoms within a certain range & 1.0 \\
Cells & Show the unit cell & off \\
Axes & Show the axes & off \\
Structure & Show the panel of \textit{Analysis of Structure} & off \\
Dynamics & Show the panel of \textit{Structural Change} & off \\
Net charge & Show Mulliken charge / Mulliken charge-C (with colors) & off \\
Spin & Show spin / spin-C (with colors) / spin-V (with vectors) & off \\
Force & Show atomic forces & off \\
Velocity & Show atomic velocity & off \\
Rot on x/y/z & Rotate the structure on x/y/z axis by selecting an angle & 0 \\
Save & Saving files: (1) a XYZ file, (2) a CIF file,                      & - \\
     & (3) an OpenMX input file (.dat) in a Cartesian coordinate system,  & \\
     & (4) an OpenMX input file (.dat) in a fractional coordinate system. & \\
Examples & Select an existing data to show crystalline or molecular structure. & - \\ \hline
\end{tabular}
\end{table}

\subsubsection{Parsing CIF files}
The original program, CIF2CELL, for parsing CIF files is written in python \cite{CIF} by using routines in the library PyCIFRW \cite{PyCIFRW}. We convert this Python code to a Javascript version, i.e. cif.js, and build it as a CIF parsing function to connect with the Structure Viewer for parsing CIF files. After a user drags and drops a CIF file to the Structure Viewer, the Structure Viewer will (1) analyze the CIF file by using the CIF parsing function, (2) get the atomic coordinates, cell vectors and corresponding symmetry, (3) send structural data to plot the structure, and (4) select a conventional cell or a primitive cell to show. After showing the structure, users can start to operate the structure in the Structure Viewer.

\subsection{Cube 3D viewer}
The Cube 3D Viewer is based on WebGL, Three.js \cite{ThreeJS}, DAT.GUI (\url{https://github.com/dataarts/dat.gui}), marching cube \cite{MarchingCube}, marching tetrahedra \cite{Tetrahedra} and surface net \cite{SurfaceNets} libraries to show 3D isosurface images in the Gaussian Cube format \cite{Gaussian} such as electron density, difference electron density taken from superposition of atomic densities of constituent elements and molecular orbitals. The Three.js and isosurface libraries are released under the MIT license and DAT.GUI is released under the Apache License 2.0. Originally, the libraries of isosurface algorithms come from Paul Bourke's work \cite{PaulBourke}. And, Mikola Lysenko transformed the codes to Javascript version and connected with Three.js to show isosurface images in a web browser \cite{MikolaLysenko}. We followed his demo for drawing isosurfaces and modified the source code to transform each small cubic cell to one of seven lattice systems for constructing crystal isosurface images.

In order to show an isosurface image, users can drag and drop a cube file to the OpenMX Structure Viewer window or the Cube 3D Viewer window as shown in Fig. 3. After reading a cube data in the 'cube.js' function and declaring two float arrays to store positive and negative values with a default isovalue in 'read3ddata.js' function, these two positive and negative data sets are sent to one of the isosurface algorithms for producing facets or edges of two isosurfaces with different colors, respectively. In the last part, by Three.js, atoms, the unit cell, and the isosurface can be presented on the Cube 3D Viewer. In addition, a default isovalue is automatically determined by a largest absolute value divided by 200. And, the isovalue can be adjusted or changed continuously with a delay time by using the drop-down menu. In Table \ref{table:t4}, the performance of the Cube 3D Viewer is given for reference. The speed of compression depends on the size of a cube data. The rendering time depends on the total number of grids and an isovalue because it is related to the numbers of vertices and edges created by isosurface algorithms and plotted by the Three.js library. In the case of surface nets, the numbers of vertices and edges required for creating an isosurface image are usually less than those in the marching cubes algorithm and marching tetrahedra algorithm. The use of memory and performance are better than other two cases. Moreover, the drop-down menu provides several options of display styles for users to choose. These options are listed in Table \ref{table:t5}. Among them, four useful options are described as below:

\begin{enumerate}
\item A supercell can be chosen to generate larger and continuous isosurface images. However, it should be noted that plotting isosurface images will become heavier if there are many facets needed to plot in the original crystal lattice. Drawing facets or edges is the most time-consuming part in the Cube 3D Viewer.
\item Edges of isosurface images can be shown by selecting 'Edges'. After edges are shown, facets will be hidden at the same time, and vice versa.
\item Atoms can be presented in the unit cell by checking the 'atoms' option. But, if atoms are behind facets, atoms will become hidden.
\item Evolution of isosurface can be performed by checking the 'start' option. Both of the isovalue and isosurface will be changed gradually with a positive or negative increment after a time delay, where a positive or negative increment can be controlled by checking the 'direction' option.
\end{enumerate}

\begin{table}[h!]
\footnotesize
\caption{Performance of creating and rendering an isosurface of fullerene-C\textsubscript{60} with 3 different meshes.} \label{table:t4} 
\begin{tabular}{crcrrrrrr}
\hline
\multirow{2}{*}{\begin{tabular}[c]{@{}c@{}}System\\ (Grids)\end{tabular}} & \multicolumn{1}{c}{\multirow{2}{*}{\begin{tabular}[c]{@{}c@{}}File compressing\\ and\\ decompressing {[}s{]}\end{tabular}}} & \multirow{2}{*}{Isovalue} & \multicolumn{3}{c}{Creating meshes and rendering {[}s{]}} & \multicolumn{3}{c}{JavaScript Memory {[}MB{]}} \\ \cline{4-9} 
 & \multicolumn{1}{c}{} &  & \multicolumn{1}{c}{\begin{tabular}[c]{@{}c@{}}Marching\\ cubes\end{tabular}} & \multicolumn{1}{c}{\begin{tabular}[c]{@{}c@{}}Marching\\ tetrahedra\end{tabular}} & \multicolumn{1}{c}{\begin{tabular}[c]{@{}c@{}}Surface\\ nets\end{tabular}} & \multicolumn{1}{c}{\begin{tabular}[c]{@{}c@{}}Marching\\ cubes\end{tabular}} & \multicolumn{1}{c}{\begin{tabular}[c]{@{}c@{}}Marching\\ tetrahedra\end{tabular}} & \multicolumn{1}{c}{\begin{tabular}[c]{@{}c@{}}Surface\\ nets\end{tabular}} \\ \hline
C60 (84$\times$84$\times$84) & 1.955 / 0.411 & 0.0284 & 1.097 & 2.305 & 0.957 & 213.9 & 597.1 & 103.9 \\
C60 (125$\times$125$\times$125) & 6.369 / 1.354 & 0.0349 & 4.121 & 7.712 & 3.668 & 357.4 & 897.8 & 263.6 \\
C60 (150$\times$150$\times$150) & 11.956 / 2.374 & 0.0349 & 7.942 & 16.859 & 7.111 & 750.6 & 1272.8 & 465.0 \\ \hline
\end{tabular}
\begin{tablenotes}
  \footnotesize
  \item \textit{Notes:} Tests for the Cube 3D Viewer were performed in a Chrome browser in Linux Ubuntu 16.04 with a 4.2 GHz Intel Core i7-7700K Processor and 32 GB DDR4-2400 MHz Memory.
\end{tablenotes}
\end{table}

\begin{figure}[h]
	\centering
	\includegraphics[scale=1.3]{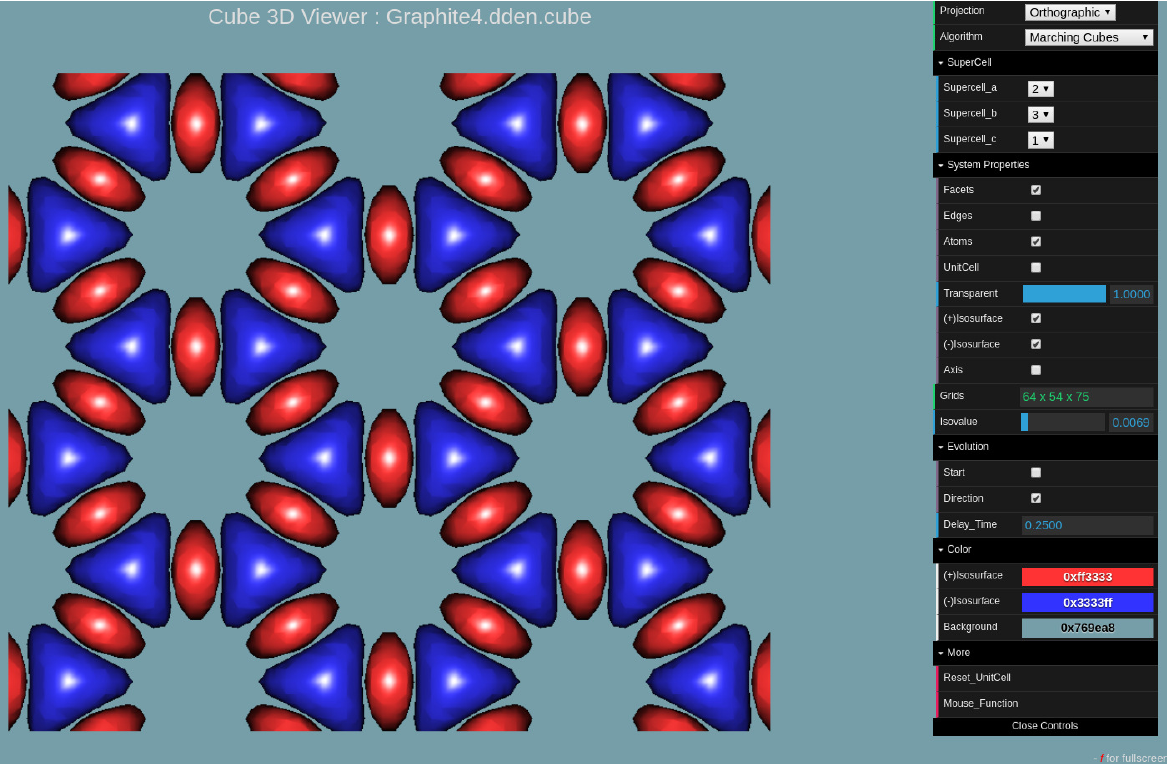} \label{figure:f3} 
	\caption{An isosurface of difference electron density of graphene is shown by the Cube 3D Viewer. The drop-down menu is \textcolor{red}{i}n the top-right corner.}
\end{figure}

\begin{table}[h!]
\caption{The options of display styles in the Cube 3D Viewer} \label{table:t5} 
\begin{tabular}{lll}
\hline
\multicolumn{1}{c}{\textbf{Options}} & \multicolumn{1}{c}{\textbf{Description}} & \multicolumn{1}{c}{\textbf{Default Value}} \\ \hline
Projection & Orthographic or Perspective projection & Orthographic \\
Algorithm & \begin{tabular}[c]{@{}l@{}}3D isosurface algorithm:\\ Marching Cubes / Marching Tetrahedra / Naive Surface Nets\end{tabular} & Marching Cubes \\
Supercell & Setup the size of supercell a $\times$ b $\times$ c & 1 $\times$ 1 $\times$ 1 \\ \hline
\multicolumn{3}{l}{\textbf{\textit{System Properties:}}} \\
Facets & Show facets of isosurface & On \\
Edges & Show lines of isosurface & Off \\
Atoms & Show atoms & Off \\
Unit Cell & Show the unit cell & On \\
Opacity & \begin{tabular}[c]{@{}l@{}}The percentage of allowing the transmission of light\\ through a material (Range = 0.0$\sim$1.0.)\end{tabular} & 1.0 \\
(+) Isosurface & Show the positive isosurface & On \\
(-) Isosurface & Show the negative isosurface & On \\
Axis & Show axes & Off \\
Isovalue & Setup an isovalue to show corresponding isosurface & $|Maximum|/200$\\ \hline
\multicolumn{3}{l}{\textbf{\textit{Evolution:}}} \\
Start & Start to change isovalue continuously & Off \\
Direction & Increase a positive (on) or negative (off) isovalue continuously & On \\
Delay\_Time & Change in isovalue per delay time (Unit = second) & 0.25 (s)\\ \hline
\end{tabular}
\end{table}

\subsubsection{Reading a cube file to get 3D-data}
In order to show a 3D isosurface, analysis functions in Javascript for reading the cube format and storing a cube data in two float arrays are prepared first. Once the cube data is obtained, an isosurface with a default isovalue will be plotted by using one of the isosurface algorithms. However, when the cube data is larger than 5 megabytes (MB), web browsers usually do not allow us to transfer the data from the OpenMX Structure Viewer window to the Cube 3D Viewer window because the localStorage in a web browser is usually limited to 5 MB. The dropped data will be compressed to strings first by using a MIT-licensed 'lz-string' compressed library (\url{https://github.com/pieroxy/lz-string}) and then these compressed strings are transferred to the Cube 3D Viewer window. In case of a large-sized cube data, the step of compressing the data can be a bottleneck, leading to time lag in the response. If users want to analyze a huge cube data, they have to recognize this issue because it will cause the computer to lag.

\subsubsection{Transformation of cell vectors}
Originally, a 3D isosurface image by using marching cubes algorithm, marching tetrahedra algorithm, or surface nets is composed of many small cubic cells. A dropped cube file is first analyzed by assuming that the lattice system is cubic, regardless of the actual lattice, to calculate facets using one of the isosurface algorithms. In order to apply isosurface algorithms for all kinds of lattice systems, Affine transformation is used to transform basis vectors of each small cubic cell to corresponding lattice vectors. After all of small cells are transformed, vertices and edges in small cells can be established and plotted by the Three.js library to visualize an isosurface image.

\subsection{Band Structure Viewer}
The Band Structure Viewer is developed for showing a band structure quickly by dragging-and-dropping a Band file in the end of OpenMX's band structure calculation as shown in Fig. 4. It supports a function for zooming in/out to a certain {band structure range} by controlling the mouse wheel or buttons. The drop-down menu provides options for scaling a band structure and setting display styles.

\begin{figure}[h!]
	\centering
	\includegraphics[scale=0.8]{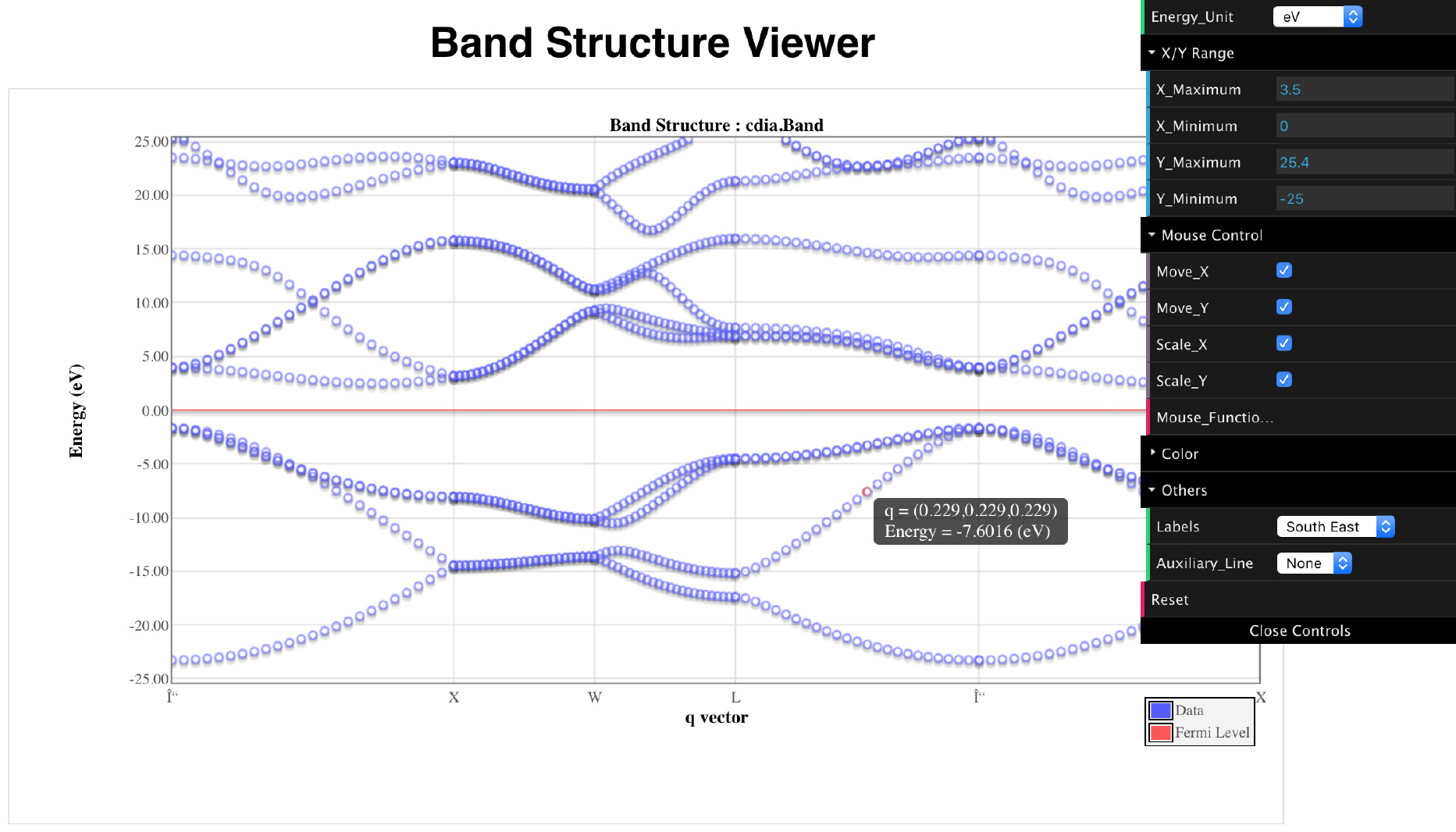} \label{figure:f4} 
	\caption{The band structure of diamond in the Band Structure Viewer, where the Fermi level is set to zero.}
\end{figure}

\section{Conclusions}
The OpenMX Viewer is an interactive web-based crystalline and molecular structure viewer for users to: (1) visualize crystal structures, isosurfaces and band strucutres; (2) analyze static/dynamic data of OpenMX; and, (3) save structural information in a XYZ format, a CIF format, or an OpenMX input format. It provides three kinds of viewers, Structure Viewer, Cube 3D Viewer and Band Structure Viewer, to check various properties by dragging-and-dropping files conveniently in a compatible web browser without any installations, upgrades, updates and terminal commands. The Structure Viewer is capable of showing crystal structures, measuring bond lengths, bond angles, and a dihedral angle, and playing an animation of molecular dynamics simulations step-by-step or continuously. The Cube 3D Viewer can show an isosurface of electron density or molecular orbitals by reading a cube file. The Band Structure Viewer can be utilized by dragging-and-dropping a Band file to show a band structure directly in a web browser. In addition to OpenMX input/output formats, XYZ, CIF, and cube formats are accepted as an input file by the OpenMX Viewer. Users can always use the newest version of the OpenMX Viewer by accessing the website. In the near future, we plan to release the OpenMX Structure Builder for users to construct or modify their structures.

\section*{Acknowledgments}
This paper is partly based on results obtained from a project commissioned by the New Energy and Industrial Technology Development Organization of Japan (NEDO) Grant (P16010). We thank Dr. Adam D. Sproson (Atmosphere and Ocean Research Institute, Univeristy of Tokyo) for proofreading the manuscript and reviewers for giving valuable comments and suggestions.

%\normalsize
\section*{References}
% -start- references

% - end - references

\bibliography{mybibfile}

\end{document}